\documentclass[aps,prl,groupedaddress]{revtex4-2}

\usepackage{graphicx}%
\usepackage{amsmath,amssymb,amsfonts}%
\usepackage{amsthm}%
\usepackage{xcolor}%

\begin{document}


\title{X(2370) as the pseudoscalar glueball: Another clue from a constituent approach}


\author{F. Buisseret}
\email[]{fabien.buisseret@umons.ac.be}

\affiliation{Service de Physique Nucl\'eaire et Subnucl\'eaire, Universit\'e de Mons, 20 Place du Parc, Mons, 7000, Belgium. \\ Laboratoire Forme et Fonctionnement Humain (FFH), Haute Ecole Louvain en Hainaut, Rue Trieu Kaisin 136, Montignies-sur-Sambre, 6061, Belgium}


\date{\today}

\begin{abstract}
The pseudoscalar state X(2370) is the clearest glueball candidate experimentally identified so far. From a phenomenological point of view, constituent approaches, in which the lightest glueballs are modelled as bound states of two transverse constituent gluons linked by an adjoint flux tube, reach a good agreement with quenched lattice QCD pure gauge spectrum. Based on the assumption that the total decay width of glueballs or light unflavoured mesons is proportional to the energy stored in the flux tube, a model is developed that linearly correlates the hadron decay width to its  mass. The model is first fitted on light unflavoured mesons and then extended to glueballs. Both the mass and decay width of the X(2370) are compatible with the results of the constituent approach. Other experimental glueball candidates in the scalar and tensor channels are also reviewed to  show the ability of the developed approach to identify glueballs: The f$_2$(2150) and f$_2$(2300) states appear as promising tensor glueball candidates. 
\end{abstract}


\maketitle

The experimental detection of glueballs has arguably reached a new milestone with the discovery of the X(2370) hadron \cite{BESIII:2023wfi,BESIII:2026rzt}. This flavour-singlet, $0^{-+}$, state with mass $2359^{+13}_{-14}$ MeV and width 170$^{+44}_{-29}$ MeV has features consistent with what is expected from the pseudoscalar glueball, such as narrow decay width and suppression of radiative decays to $\omega$ and $\phi$ \cite{BESIII:2026mvn}. The X(2370) has moreover a mass near the lower limit of benchmark quenched SU(3) lattice QCD computations: 2590$\pm$170 MeV \cite{Morningstar:1999rf}, 2560$\pm$155 MeV \cite{Chen:2005mg}, 2561$\pm$40 MeV \cite{cite-key}. It is even closer to the mass reported in \cite{Meyer:2004gx,Meyer:2004jc}, that is 2250$\pm$170 MeV. 

On the theoretical side, the study of Yang-Mills spectrum within effective approaches has motivated a great number of studies, see e.g the reviews \cite{Mathieu:2008me,Llanes-Estrada:2021evz}. Among these approaches, constituent models, assuming that glueballs are bound states of transverse massless gluons, give a satisfactory qualitative \cite{Boulanger:2008aj} and quantitative \cite{Mathieu:2008bf,Chevalier:2025xed} description of pure gauge lattice data using a spinless Salpeter Hamiltonian with funnel potential and helicity states, see the pioneering works \cite{Jacob:1959at,Wick:1962zz}. In the $C=+$ sector for example, the simplest glueball Hamiltonian is the two-body Hamiltonian 
\begin{equation}\label{ham0}
	H=2\sqrt{\vec p^{\, 2}}+a\, r-\frac{3\alpha_s}{r},
\end{equation}
where the orbital and spin parts of the wave functions are given by helicity states for two transverse spin 1 particles, see \cite{Mathieu:2008bf} for explicit expressions. Note that, although the other channels are more involved, the $0^{-+}$ glueball helicity state is a pure $^3P_0$ state.  In the above Hamiltonian, $a\, r$ is the potential energy of the flux tube or QCD string, $a$ is the adjoint string tension and $\alpha_S$ is the strong coupling constant. Note that $a= \frac{9}{4}\sigma$ according to the Casimir scaling hypothesis (see e.g. \cite{Bali:2000un} and references therein), with $\sigma$ the fundamental string tension. The Casimir scaling hypothesis has recently proven to reproduce the observed trends of the glueball masses versus $N$ in SU($N$) lattice gauge computations~\cite{Buisseret:2025qge}.

Using the relativistic virial theorem \cite[Eq. (8)]{PhysRevLett.64.2733}, one can express the mass spectrum of $H$ as
\begin{equation}\label{Mass}
	M=	\left\langle H \right\rangle = 2 E_{FT},
\end{equation}
with 
\begin{equation}\label{EFT}
	E_{FT}=a \left\langle r \right\rangle 
\end{equation}
the energy stored in the adjoint flux tube. One could wonder whether flux tube width should be present in $E_{FT}$ but it has been shown in \cite{Semay:2004br} that, within the Casimir scaling hypothesis, the flux tube width stays constant no matter its representation, fundamental or adjoint. Relativistic effects in the flux-tube energy are neglected in the present approach. Their inclusion is discussed for mesons in \cite{PhysRevC.76.025206} and more specifically for glueballs in \cite{Buisseret:2009yv}. For our purpose, it is important to outline two results of these last references. First, a constituent approach based on an unperturbed Hamiltonian of the form (\ref{ham0}) may accurately reproduce light and heavy meson mass spectra provided relativistic, spin-dependent, corrections are introduced in perturbation \cite{PhysRevC.76.025206}. Second, such relativistic effects do not modify the qualitative behaviour of the mass spectrum in the $C=+$ glueball sector, and can be absorbed in a rescaling of the string tension and of the strong coupling constant in (\ref{ham0}). The crucial ingredient is actually the transverse nature of the constituent gluons. 

\begin{figure*}[t]
	\centering
	\includegraphics[width=0.8\textwidth]{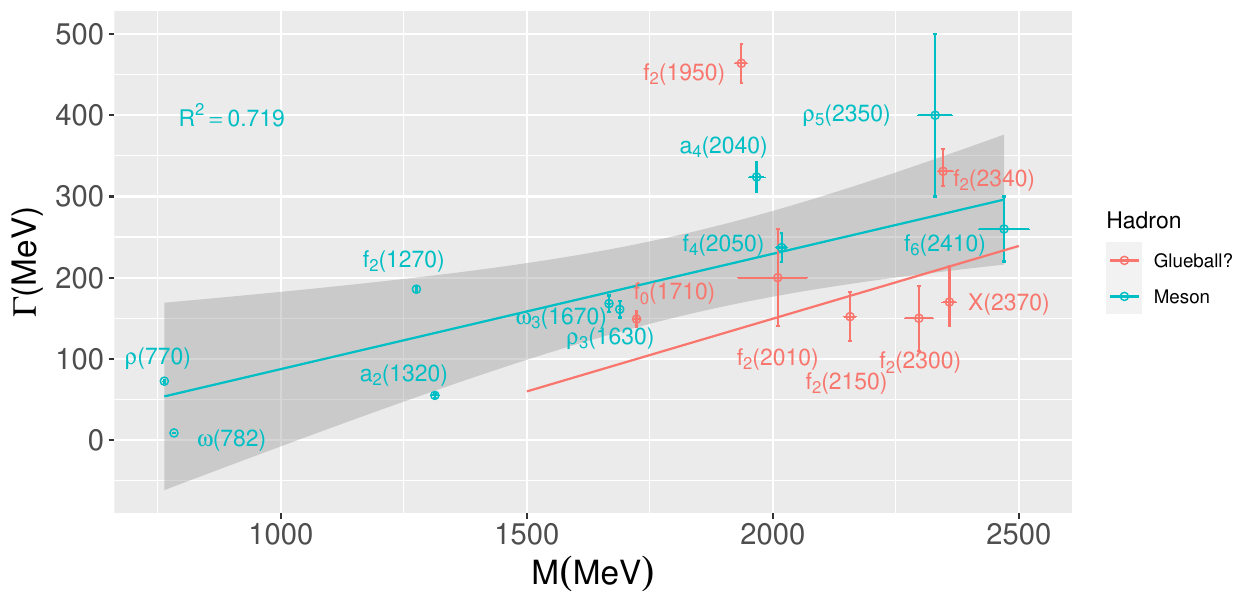}
	\caption{Total decay width $\Gamma$ vs mass $M$ for the orbital excitations of the $n=0$ unflavoured mesons (blue points$+$error bars); data are taken from the Particle Data Group \cite{ParticleDataGroup:2026oor}. The fit (\ref{fit_mes}) of model (\ref{Gamma_mes}) is displayed (blue line) with its 95\% confidence interval. Model (\ref{Gamma_glu}), using the latter fit, is shown (red line) together with several glueball candidates (red points$+$error bars).  }\label{fig1}
\end{figure*}

In order to assess the compatibility of the X(2370) with a glueball state within the proposed approach, a computation of the glueball decay width is needed. This will be done by developing further the proposal of \cite{Abreu:2005uw}. Let us start with the case of light mesons, seen as bound states of a constituent quark and a constituent antiquark linked by a chromoelectric gluonic field, i.e. the flux tube. Within this flux tube, a particle-antiparticle pair can be created from the vacuum thanks to the Schwinger effect \cite{Schwinger:1951nm} and lead to a decay by flux-tube breaking. It is known that such a mechanism can successfully describe the decay widths of mesons, and even baryons \cite{Kokoski}. In light mesons, decays induced by flux-tube breaking should be dominant over short-range effects. The larger $E_{FT}$, the more probable the creation of a $q\bar q$ pair. We therefore propose, along the lines of \cite{Abreu:2005uw}, that the total decay width $\Gamma_{q\bar q}$ is estimated by $\Gamma_{q\bar q}=\gamma\left(E_{FT}-E_0\right)$ with $\gamma=\frac{d\Gamma}{dE_{FT}}$ a constant. Using (\ref{Mass}), one reaches the model
\begin{equation}\label{Gamma_mes}
	\Gamma_{q\bar q}(M_{q\bar q})=\gamma\left(\frac{M_{q\bar q}}{2}-E_0\right),
\end{equation}
where $E_0$ is the minimal energy needed to create a $q\bar q$ pair. As shown in Fig. \ref{fig1}, fitted formula (\ref{Gamma_mes}) can successfully  ($R^2=0.719$) reproduce the trend observed on the $n=0$ unflavoured light meson states, the states for which our model is mostly applicable a priori. One obtains $\Gamma_{q\bar q}=(0.1792\pm 0.037) M_{q\bar q}-104\pm 63$~MeV, or more explicitly 
\begin{equation}\label{fit_mes}
	\gamma=0.3584\pm 0.074   ,\quad E_0=580\pm 471\ {\rm MeV}.
\end{equation}
$E_0$ can be interpreted as twice the energy of a constituent quark with mass $290$ MeV, in coherence with generally accepted values in the range 150-300~MeV \cite{ROBERTS2021103883}. It has to be said that constituent approaches, such as the one developed here, are not intended to provide high-precision results (unless a too large number of ad hoc parameters is used) but rather to give trends, even analytical ones, able to guide the interpretation of experimental or lattice data. 

In the glueball case, breaking an adjoint flux tube demands the creation of two adjoint quasiparticles, called gluelumps in lattice QCD \cite{DEFORCRAND2000280}. In our scenario, these are two $q\bar q$ adjoint pairs. Hence, one can estimate that the minimal energy needed to break the adjoint string is that of 2 $q\bar q$ pairs. This leads to the following model for the two-gluon glueball decay width:
\begin{equation}\label{Gamma_glu}
	\Gamma_{gg}(M_{gg})=\gamma\left(\frac{M_{gg}}{2}-2E_0\right), 
\end{equation}
suggesting that a glueball will always be narrower than a light meson with the same mass, i.e. $\Gamma_{gg}(M)<\Gamma_{q\bar q}(M)$. 

The expected trend for $\Gamma_{gg}$ vs $M_{gg}$ is displayed in Fig. \ref{fig1} as well as the X(2370). Other glueball candidates are displayed for completeness: The $f_0(1710)$ that is expected to have a dominant scalar glueball component (see the review \cite{Llanes-Estrada:2021evz} for a detailed discussion), and $f_2$ states whose status as dominantly tensor glueball is still under debate \cite[Spectroscopy of Light Meson Resonances]{ParticleDataGroup:2026oor}.The X(2370) lies outside the 95\% confidence interval of the meson model and is rather compatible with our glueball model up to the error bars. The $f_0(1710)$ seems equally close to the meson and glueball models after graphical inspection. The $f_2(1950)$ is too wide regarding (\ref{Gamma_glu}), while the $f_2(2010)$ and $f_2(2340)$ fall in the meson sector. The $f_2(2150)$ and $f_2(2300)$ are favoured tensor glueball candidates within the present approach. We point our that the $f_2(1950)$ has been identified as a tensor glueball by chiral models \cite{Vereijken:2023jor}, and that chiral effects are not included in this work.

\begin{table*}[t]
	\caption{``Experiment" columns: Several experimental glueball candidates including the X(2370). ``Model" columns: Masses computed the the model (\ref{ham0}) with $a=\frac{9}{4}\, 0.185$  GeV$^2$, $\alpha_S=0.45$, and  two-transverse gluons helicity states, $M_{gg}$. Decay widths, $\Gamma_{gg}(M_{exp})$ and $\Gamma_{gg}(M_{gg})$, are computed with the best values of the fit defined by Eqs. (\ref{Gamma_glu}) and (\ref{fit_mes}); the error reported comes from the largest experimental error on the hadron mass. All numbers are given in MeV}\label{tab1}
	\begin{tabular*}{\textwidth}{@{\extracolsep\fill}c|cll|cll}
	\hline
		& \multicolumn{3}{@{}c@{}|}{Experiment} & \multicolumn{3}{@{}c@{}}{Model} \\
		$J^{PC}$ & State & $M_{exp}$  & $\Gamma_{exp}$ & $M_{gg}$ \cite{Chevalier:2025xed} & $\Gamma_{gg}(M_{exp})$ & $\Gamma_{gg}(M_{gg})$ \\
\hline
		$0^{-+}$ & X(2370) \cite{BESIII:2026rzt}& $2359^{+13}_{-14}$  &170$^{+44}_{-29}$  &  2216 & 215$\pm$3  & 189\\
		$0^{++}$ & $f_0(1710)$ \cite{ParticleDataGroup:2026oor}& $1723^{+7}_{-6}$ & 149$\pm$10 & 1769 &101$\pm$1  & 109 \\
		$2^{++}$ & $f_2(1950)$\cite{ParticleDataGroup:2026oor} & $1936\pm12$& 464$\pm$24 & 2279 & 139$\pm$2  & 200\\
		& $f_2(2010)$ \cite{ParticleDataGroup:2026oor} & 2010$^{+60}_{-80}$ & 200$\pm$60& 2279 & 152$\pm$14  & 200\\ 
		& $f_2(2150)$ \cite{ParticleDataGroup:2026oor} & 2157$\pm$12 & 152$\pm$30& 2279 & 179$\pm$2  &200 \\ 
		& $f_2(2300)$ \cite{ParticleDataGroup:2026oor} & 2297$\pm$28 & 150$\pm$40& 2279 & 202$\pm$5 &200 \\ 
		& $f_2(2340)$ \cite{ParticleDataGroup:2026oor} & 2346$^{+21}_{-10}$ & 331$^{+27}_{-18}$ & 2279 & 212$\pm$4 & 200\\ 
\hline
	\end{tabular*}
	
\end{table*}

The mass spectrum of Hamiltonian (\ref{ham0}) and its three-body generalization (leading to $C=-$ glueballs) has been computed in \cite{Chevalier:2025xed} within a unified framework allowing to handle two- and three-transverse gluon helicity states. The fundamental tension was set to $\sigma=0.185$  GeV$^2$, that is the value leading to the best agreement with experimental data in the light and heavy meson case \cite{PhysRevC.76.025206}, and $\alpha_S=0.45$ was fitted on the $C=+$ quenched SU(3) lattice QCD spectrum, leading also to a good agreement in the $C=-$ sector \cite{Chevalier:2025xed}. Relevant values are reported in Table~\ref{tab1}. 

The model predicts a pseudoscalar mass in agreement with the X(2370) up to 6\% and a decay width that falls within the experimental error bars, adding evidence of the identification of this state as the pseudoscalar glueball. The $f_0(1710)$ has predicted mass and decay width close to our estimates also; its larger experimental decay width could be induced by mesonic $n\bar n$ or $s\bar s$ components besides $gg$. Moreover, the $f_2(2300)$ is the favoured tensor glueball candidate according to the proposed model.

\end{document}